\documentclass[11pt]{article}
\usepackage{moriond,epsfig}

\def\be{\begin{equation}}
\def\ee{\end{equation}}
\def\bea{\begin{eqnarray}}
\def\eea{\end{eqnarray}}

\begin{document}
\vspace*{4cm}
\title{THE MASS MISSING PROBLEM IN CLUSTERS: DARK MATTER OR 
MODIFIED DYNAMICS?}

\author{ E. POINTECOUTEAU }

\address{CESR, 9 av. du colonel Roche, BP4346, 31028 Toulouse, France}

\maketitle\abstracts{ The widely accepted dark matter hypothesis
  offers a seductive solution to missing mass problems (galaxies,
  clusters of galaxies, gravitational collapse in structure
  formation,...). However the physical nature of the Dark Matter
  itself is still unknown. Alternatively, it has been proposed that
  apparent dynamical evidence of dark matter is due to a modification
  of Newton's law of gravitation. Here we revisit the Modified
  Newtonian Dynamics (MOND) theories at the scale of galaxy
  clusters. Using hydrodynamical simulations, we derived quantities
  such as the density and the temperature of the ICM. We compared
  those MOND simulated predictions to high quality X-ray density and
  temperature profiles observed down to $\sim$0.5 the virial ra-
  dius. If the density profiles seems in acceptable agreement, the
  simulated temperature show a constant increase with the radius
  whereas the observed profiles show a flat to a mild decrease shape
  down to $\sim$0.5 $R_{virial}$. We also computed the dynamical MOND mass
  for 8 X-ray clusters observed with XMM-Newton. If the MOND mass
  helps to lower the discrepancy with the baryonic mass by $\sim$20\%,
  still $\sim$80\% of the mass in clusters is unaccounted by
  baryons. In order to solve this problem and to reconcile MOND with
  clusters observations, we investigated the possibility of an added
  dark baryonic component. We assumed a component of massive neutrinos
  to fill the remaining discrepancy between the observed MOND
  dynamical mass and the baryonic mass. This led us to derive a tied
  observational lower limit for the neutrino mass, $m_\nu=1.06$~eV. }

\section{Introduction }

The missing mass problem in clusters of galaxies arises from the
comparison of the observed baryonic mass with the observed dynamic
mass. The baryonic mass is mainly due to the hot intracluster gas that
is well observed in X-rays via its free-free emission. The current
status of the observed gas fraction in clusters gives a fairly well
constrained value of about 12\% [see 10, 2, for instance]. Taking into
account the stellar mass, this makes the discrepancy between the
observed dynamic mass and the observed baryonic mass larger than a
factor of 7.

\begin{figure}[!t]
\begin{center}
\includegraphics[width=8cm]{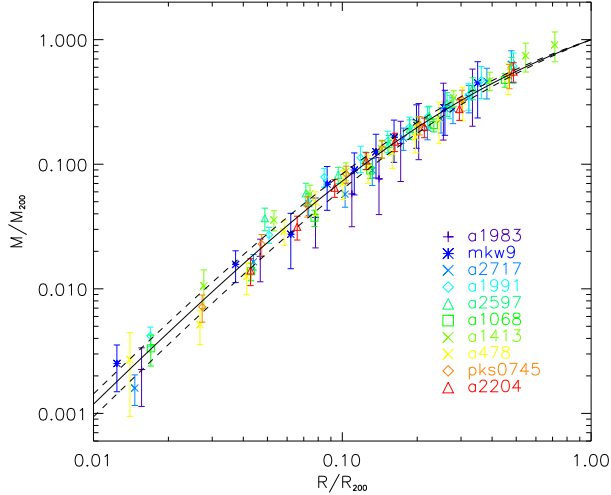}
\end{center}
\caption{Scaled mass profiles of ten nearby clusters. The mass is
  scaled to $M_v$ , and the radius to $R_v$ , both values being
  derived from the best fitting NFW model. The solid black line
  corresponds to the mean scaled NFW profile and the two dashed lines
  are the associated standard deviation. Figure from [16].}
\end{figure}

We investigate in the following two hypothesis to solve this problem,
the Dark Matter paradigm and the Modified Newtonian Dynamics paradigm.

\section{The Dark side of the matter}
The dark matter (DM hereafter) hypothesis appears to provide a
seductive explanation of the mass missing problem. A new component of
non-baryonic matter, insensitive to all interactions but gravitation,
is introduced to fill the gap between the baryonic matter and the
binding mass.

[16] have measured the total mass profile over a sample of ten
clusters from $0.01$~R$_v$ up to $0.5$~R$_{200}$ using XMM-Newton
observations. Their sample has an excellent temperature coverage and
covers an order of magnitude in mass from $M_v = 1.2\times 10^{14}~$~M$_\odot$
to $1.2\times 10^{15}$~M$_\odot$.

They have found that the NFW profile is a good representation of the
ten observed mass profiles, and that in all cases the isothermal
sphere model (i.e a profile with a core) is rejected at high
confidence. In other words, they confirm the cusped nature of the Dark
Matter profile, as predicted by CDM simulations of hierarchical
structure formation, over the temperature/mass range of the
sample. The mass profile shape is close to universal, again as
predicted, with a dispersion of less than 15\% at $0.1$~R$_v$ in the
scaled mass profiles. The shape is quantitatively consistent with
theoretical predictions. The variation of the observed concentration
parameters with mass is in line with the predictions, taking into
account the measurement errors and the expected intrinsic
scatter. Taken together,these results provide strong evidence in
favour of the Cold Dark Matter cosmological scenario, and show that
the physics of the Dark Matter collapse is well understood. The ten
scaled profiles are presented on Fig. 1.

However, while cosmological evidence is accumulating in favour of this
scenario (see for instance [7, 22, 23]), it is disconcerting that the
nature of the non-baryonic dark matter is completely unknown. Of
course there are many candidates of varying degrees of detectability
and plausibility (e.g review by [4]).

\section{Living in a MOND world}
As an alternative to dark matter scenarios, [13] proposed a
modification of the Newtonian dynamics effective at galactic and
extra-galactic scales. This modified Newtonian dynamics (MOND
hereafter) has been notably successful in explaining the discrepancy
between rotation and luminosity curves in spiral galaxies [11, 12],
and claims other phenomenological successes (for a full review on MOND
see [19]). The discrepancy between the baryonic mass and the dynamical
mass in clusters of galaxies is perhaps foremost among the issues that
MOND has yet to convincingly address.
 
The first confrontation of X-ray observations of clusters with MOND
[9] emphasised the difficulties faced by MOND in passing the cluster
test. The problem was revisited by [17, 18] and ended in a remaining
discrepancy of a factor of 2-3 between the baryonic and the MOND
masses. More recently, [1] discussed observational evidence for three
clusters for which the observed discrepancy is about 1-5 within 1 Mpc
and is boosted to a factor $\sim$10 within the central 200 kpc, further
weakening the reliability of the MOND paradigm. However, [20]
responded with an update of his earlier work, introducing an added ad
hoc dark component at the cluster centre to reduce the discrepancy to
only a factor of 1-3 overall in the cluster.

Some other tests have also been carried out using gravitational
lensing data. They have also pointed out the difficulties faced by
MOND at the cluster scale [8, 5].

\section{Revisiting MOND at galaxy clusters scale}

Recently, [15] and [14] have revisited the MOND paradigm at the
cluster scale. In the following we present a synthesis of their
results.

\subsection{Structure formation in MOND}
[14] use a one dimensional hydrodynamical code to study the evolution
of spherically symmetric perturbations in the framework of Modified
Newtonian Dynamics (MOND). The code used evolves spherical gaseous
shells in an expanding Universe by employing a MOND-type relation-
ship between the fluctuations in the density field and the
gravitational force, $g$. They focus on the evolution of initial density
perturbations of the form $\delta_i \sim r^{-s}_i$ for $0 < s < 3$. A
shell is initially cold and remains so until it encounters the shock
formed by the earlier collapse of shells nearer to the centre. During
the early epochs g is sufficiently large and shells move according to
Newtonian gravity. As the physical size of the perturbation increases
with time, g gets smaller and the evolution eventually becomes MOND
dominated. However, the density in the inner collapsed regions is
large enough that they re-enter the Newtonian regime. The evolved gas
temperature and density profiles tend to a universal form that is
independent of the the slope, s, and of the initial amplitude.

The results from the simulations, the temperature and density profiles
of MOND, were confronted with recent X-ray observations of nearby
galaxy clusters by XMM-Newton [16] and Chandra [24]. Among the
different MOND runs, the minimum value number density at 500 kpc in
MOND is $0.07$~cm$^{-3}$ . This is substantially higher that the
maximum observed number density of $0.0026$~cm$^{-3}$. More
importantly, the temperature profile, $\sim r^{0.5}$ in MOND, is in clear
contrast to the observed profiles, which either flatten or show a mild
decline in the outer regions.

\subsection{Observed clusters mass profiles versus MOND predictions}
From the mass profiles derived by [16] in a LCDM cosmology, [15]
computed the observed ratio of the dynamical MOND mass, $M_m$ to the
baryonic mass $M_d$ :
 
\begin{equation}
\frac{M_m}{M_b}(r) = \left[f_{gas} (1+f_\star)\sqrt{1+\left(\frac{a_0}{G}\frac{r^2}{M_d}\right)}\right]^{-1}
\end{equation}

\begin{figure}[!t]
\begin{center}
\includegraphics[width=8cm]{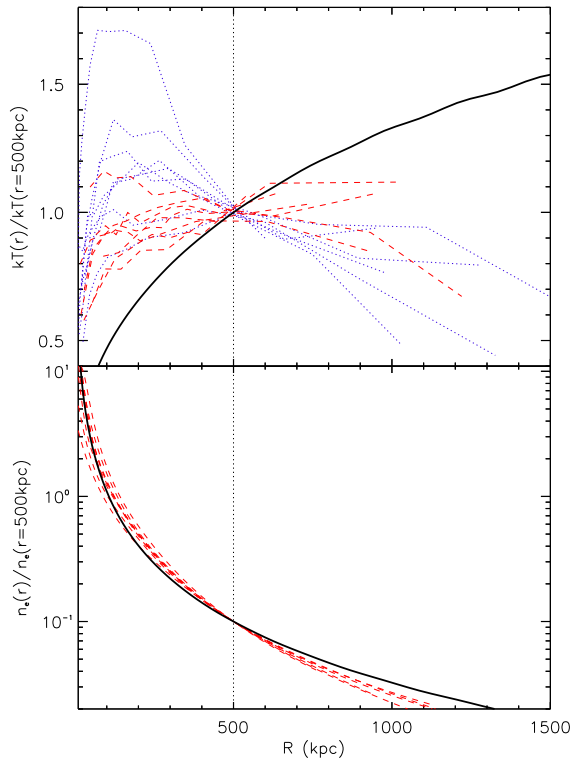}
\end{center}
\caption{MOND simulated temperature profiles (thick black
  line) compared to the individual profiles of clusters observed with
  XMM-Newton (red dashed lines) and Chandra (blue dotted
  lines). Bottom panel: MOND simulated density profiles (thick black
  line) shown with respect to the observed density profiles of nearby
  clusters by the XMM-Newton satellite (red dashed lines). In both
  panels the profiles have been normalised according to their
  respective values at $r = 500$~kpc as marked radius by dotted vertical
  line. Figure from [14].}
\end{figure}

where $f_{gas}$ is the gas fraction, $f_\star$ is the stellar fraction
(i.e the ratio of the stellar mass to the gas mass of the
cluster). $a_0 \sim 10^{-8}$cm/s$^{-2}$ is a fundamental accelaration
constant in the MOND theory (its value is derived from the analysis of
the rotation curves of galaxies in a MOND framework), $G$ is the
gravitational constant, $M_d$ is the hydrostatic mass of the cluster as
obtained from the hydrostatic equilibrium equation.

In a pure MOND universe this ratio should equate 1. Table~1 reports
the average baryon fraction values (i.e $f_b = f_{gas} (1 + f_\star
)$) and the average $M_m/M_b$ ratios for the measured clusters at
different radius. The computation were done at the following radii:
the radii corresponding to the density contrasts, with respect to the
critical density of the Universe at the cluster redshift of $\delta =
1000$ and $\delta = 15000$ (i.e $0.47 \pm 0.02$~R$_{200}$ and $0.10
\pm 0.01$~R$_{200}$ average over the eight out of ten clusters from
[16]). Those two radii mark the boundaries of the radial range
over which the observational constraints are especially well tied
down.

\begin{table}
\caption{Ratios of the dynamical mass to the MOND mass and of
  the MOND mass to the baryonic mass over 8 nearby clusters.}  
\begin{center}
\begin{tabular}{lll}
\hline
\hline
Radius & $f_b$ & $M_m/M_b$ \\
\hline
$\delta= 15000$ & $0.09 \pm 0.03$ & $10.6 \pm 3.77$ \\
$\delta = 1000$ & $ 0.13 \pm 0.02$ & $4.94 \pm 0.50$ \\
\hline
\hline
\end{tabular}
\end{center}
\end{table}

 At $0.5$~R$_v$, the ratio of the MOND mass to the baryonic mass is
 $4.94 \pm 0.50$ ($10.6 \pm 3.77$ at $\sim 0.1$~R$_v$). This is more
 than a factor of two above the value derived by [18]. The evidence is
 confirmed if we only consider the hot systems. Indeed, for clusters
 with $kT > 3.5$~keV, $M_d /M_m = 1.43 \pm 0.08$ at $\sim
 0.5$~R$_{200}$. This makes the ratio of the MOND mass to the baryonic
 mass $\sim 5.10 \pm 0.56$. Thus in all cases, within a $3 \sigma$
 (i.e. 99\% confidence) the ratio $M_m/M_b$ in a MOND cosmology will
 be greater than 3.4, making MOND unable to fully overcome the missing
 mass problem in clusters. With respect to the used sample, in a MOND
 framework, still about 80\% of the total mass of galaxy clusters is
 missing at $\sim 0.5$~R$_{200}$ . In other terms, this means that
 MOND just reduces the missing mass problem in clusters by about 20\%
 (at half the virial radius), but does not solve it.

\subsection{Filling clusters with neutrinos}
A last alternative to rescue MOND is to invoke a non-luminous
component at the centre of clusters, as suggested recently by
[20]. This author proposed massive neutrinos, aggregating at the
cluster scale, as candidates for this dark component. More recently,
[21] also called upon this neutrino hypothesis, studying formation of
structures in the relativistic MOND framework (i.e. the Bekenstein
theory [3]).  

Further assuming a constant density sphere for this dark component and
taking into account the phase space density limit for neutrinos, [20]
derived an upper limit for the neutrino density after their collapse
and accretion within structure of: $\rho_\nu \leq (4.8 \times
10^{-24})(m_\nu/2\textrm{eV})^4 (\textrm{T}_{\textrm{keV}} )^{3/2}$~kg/m$^{-3}$. We make use of this
limit on the neutrino density, and we use the spectroscopic
temperatures measured for each cluster of the previous sample between
0.1 and 0.5 R$_{200}$ (see [16]). From the eight clusters, it is
possible to compute the needed neutrino mass to equate the missing
mass at a given radius (i.e. $M_m(r)-M_b (r)$) with the contribution
of the massive neutrino to the cluster total mass. [20] hypothesis of
the neutrino accretion mainly concerned the central parts of
clusters. In our study, to explain the $\sim$80\% of missing mass in
MOND, we add to extend the radius of the neutrino sphere down to
R$_{1000}$ . The minimum neutrino mass then required is $m_\nu > 1.74
\pm 0.34$~eV . This is a strongly constraining value for the neutrino
mass, which makes the lower bound for the neutrino mass becomes
$\sim1.06$~eV, within a $2\sigma$ limit (i.e. 95\% confidence). This
is barely compatible with the cosmological constraints from combined
CMB+LSS data [22, 6].

\section{Conclusion}
We have presented here the case of two hypothesis to solve the missing
mass problem at the galaxy clusters scale: The Dark Matter and the
MOND paradigms.

The Dark Matter hypothesis remains a very secure and stable grounds as
it solve by definition the missing mass problem. However, the question
of the nature of the Dark matter is still one of the most dazzling
problem in modern astrophysics.

The MOND hypothesis, as a stand alone solution, has proven to be
unable to solve the missing mass problem at the clusters scale. An
added hot dark component is needed to rescue MOND, massive neutrinos
for instance. As the amount of hot DM needed is up to $\sim$80\% of the
clusters mass, this turns the MONDian cosmological framework more into
a mixed DM cosmology.  

The results presented to this conference are extensively detailed the
three following papers: Pointecouteau, Arnaud \& Pratt (2005) [16],
Pointecouteau \& Silk (2005) [15] and Nusser \& Pointecouteau (2006)
[14].

\section*{References}

\indent\indent[1] Aguirre A., Schaye J., Quataert E., 2001, ApJ, 561, 550 

[2] Allen S. W., et al., 2003, MNRAS, 342, 287 

[3] Bekenstein J. D., 2004, Phys. Rev. D, 70, 083509 

[4] Bertone G., Hooper D., Silk J., 2004, Phys. Rep., 405, 279 

[5] Clowe D., Gonzalez A., Markevitch M., 2004, ApJ, 604, 596 

[6] Elgarøy Ø., Lahav O., 2005, New Journal of Physics, 7, 61 

[7] Freedman W. L., et al., 2001, ApJ, 553, 47 

[8] Gavazzi R., 2002, New Astronomy Review, 46, 783 

[9] Gerbal D. et al., 1992, A\&A, 262, 395 

[10] Grego L., et al., 2001, ApJ, 552, 2 

[11] Milgrom M., 1983a, ApJ, 270, 371 

[12] Milgrom M., 1983b, ApJ, 270, 384 

[13] Milgrom M., 1983c, ApJ, 270, 365 

[14] Nusser, A., Pointecouteau, E. 2006, MNRAS, 366, 969 

[15] Pointecouteau, E., Silk, J. 2005, MNRAS, 364, 654 

[16] Pointecouteau, E., Arnaud, M., Pratt, G. W. 2005, A\&A, 435, 1 

[17] Sanders R. H., 1994, A\&A, 284, L31 

[18] Sanders R. H., 1999, ApJ, 512, L23 

[19] Sanders R. H., McGaugh S. S., 2002, ARA\&A, 40, 263 

[20] Sanders R. H., 2003, MNRAS, 342, 901 

[21] Skordis C., et al., 2006, Physical Review Letters, 96, 011301 

[22] Spergel D. N., et al., 2003, ApJS, 148, 175 

[23] Tegmark M., et al., 2004, Phys. Rev. D, 69, 103501 

[24] Vikhlinin, A., et al., 2005, ApJ, 628, 655 

\end{document}